# Creating Quantum Emitters in Hexagonal Boron Nitride Deterministically on Chip-Compatible Substrates

Xiaohui Xu[1], Zachariah O. Martin[2], Demid Sychev[2], Alexei S. Lagutchev[2], Yong Chen[2,3,4], Takashi Taniguchi[5], Kenji Watanabe[5], Vladimir M. Shalaev[2], Alexandra Boltasseva[1,2]

[1]School of Materials Engineering, Purdue University, USA

[2]School of Electrical and Computer Engineering, Purdue University, USA

[3]Department of Physics and Astronomy, Purdue University, USA

[4]Department of Physics and Astronomy, Aarhus University, Denmark

[5]National Institute for Materials Science, Japan

**Abstract**

Two-dimensional hexagonal boron nitride (hBN) that hosts bright room-temperature single-photon emitters (SPEs) is a promising material platform for quantum information applications. An important step towards the practical application of hBN is the on-demand, position-controlled generation of SPEs. Several strategies have been reported to achieve the deterministic creation of hBN SPEs. However, they either rely on a substrate nanopatterning procedure that is not compatible with integrated photonic devices or utilize a radiation source that might cause unpredictable damage to hBN and underlying substrates. Here, we report a radiation- and lithography-free route to deterministically activate hBN SPEs by nanoindentation with an atomic force microscope (AFM) tip. The method is applied to thin hBN flakes (less than 25 nm in thickness) on flat silicon-dioxide-silicon substrates that can be readily integrated into on-chip photonic devices. The achieved SPEs yields are above 30% by utilizing multiple indent sizes, and a maximum yield of 36% is demonstrated for the indent size of around 400 nm. Our results mark an important step towards the deterministic creation and integration of hBN SPEs with photonic and plasmonic on-chip devices.

## Introduction

Solid-state single photon emitters (SPEs) are receiving increasing attention in the last decade due to their critical role in quantum information technologies including secure quantum communication, quantum photonic computing and quantum sensors.[1–3] SPEs are typically composed of atomic defect structures in a solid-state host material that are suitable for integration with on-chip quantum photonic systems. Recently, two-dimensional (2D) van der Waals materials such as transition-metal dichalcogenides (TMDCs)[4] and hexagonal boron nitride (hBN)[5] have been extensively investigated due to their capability to host SPEs. For instance, various types of SPEs operating at ambient conditions have been identified in hBN, with emission ranging from ultra-violet (UV) to the near infra-red (NIR) spectral regime[6,7]. The atomic-scale thickness of 2D hBN not only enables high-efficiency light extraction, but also offers unparalleled advantages for integrating SPEs with plasmonic and photonic structures for hybrid quantum devices.

The practical integration of SPEs in 2D materials requires deterministic creation or activation of emitters at designated locations. Recent studies have reported a few methods to fabricate position-controlled SPEs in 2D materials based on strain engineering, ion/electron beam irradiation and controlled bottom-up growth. Strain/curvature-induced SPEs in hBN and TMDCs have been fabricated either by employing a nano-structured substrate (e.g., with nanopillars[8], nanocones[9], etc.[10]), or by deforming the 2D materials into a soft polymetric substrate[11]. However, both approaches have limited applications in quantum integrated photonics due to the involvement of patterned substrates or soft polymers. A similar strategy was utilized in a recent work to obtain SPEs that are not purely strain-induced in hBN, by growing hBN on dielectric nanopillars using chemical vapor deposition (CVD)[12]. Aside from the above-mentioned methods, gallium focused ion beam (FIB) and electron beam have been demonstrated to create position-controlled SPEs in hBN flakes on a flat substrate [13,14]. However, the fluorescence contamination induced by high-energy gallium ion implantation could be a potential concern in practice, while SPE activation by electron beam suffers from limited spatial precision (> 1 um).

In this work, we propose to deterministically activate room-temperature hBN SPEs by nano-indenting hBN with an AFM tip. The method is demonstrated for exfoliated, thin hBN flakes (less than 25 nm in thickness) placed on an unpatterned (flat) $SiO_2$-coated silicon substrate. AFM probes with a diamond-like coating material are used to indent hBN, without notable damage to the substrates. Our method utilizing an AFM tip is contamination-free, with no additional fluorescent contaminants introduced to hBN, in contrast to radiation- or fabrication-based processes.[8,12–14] By controlling the AFM probe displacement along the vertical direction, lateral indentation sizes ranging from ~150 nm up to 800 nm are achieved with good repeatability. The nano-indentation is followed by argon annealing to fully activate SPEs at the indented sites. Our results provide a promising route to fabricate site-controlled hBN SPEs on chip-compatible substrates and pave the way to the realization of integrated quantum photonics with hBN SPEs. The capability to controllably pattern an array of quantum emitters also opens exciting possibilities such as realizing a "quantum mirror"[15–17] and arrays of quantum photonic or hybrid sensors.

**Results and Discussion**

A schematic of the proposed nanoindentation technique utilizing AFM is shown in Figure 1a. First, hBN flakes exfoliated from high-quality hBN crystals are transferred to a $SiO_2$-coated Si substrate. After solvent cleaning, thin flakes (less than 25 nm in thickness) are selected for further experiments using an optical microscope (see Figure 1b showing an AFM image of a typical hBN flake used for demonstrating the nanoindentation effect). Typically, flakes have thicknesses around 20-25 nm and roughness below 500 pm (Figure 1), indicating an extremely clean and smooth surface. In the AFM contact mode, the cantilever coated with diamond-like carbon is brought into contact with hBN and then gets indented into the flake with further cantilever displacement along the vertical direction. To ensure effective indentation, the cantilever is held at its maximum displaced position for 2 seconds before retraction. By adjusting the maximum cantilever displacement ($\Delta z_{max}$), indents with varying lateral sizes could be obtained, as shown in Figure 1c. To explore the indent structure, we scanned one of the indents (Figure 1c) and extracted a line profile across the indented area. As illustrated in Figure 1d, the area is composed of a dip created in hBN as well as two bent hBN pieces resulting from the AFM probe impact. To demonstrate how the lateral size of indented dimples in hBN scales with the cantilever displacement, we measured all the indent sizes as a function of $\Delta z_{max}$ (Figure 1c and Figure 1e). The method used to determine lateral indent sizes in this work could be found in Section I, Supplementary Information. The lateral indent size scales almost linearly with $\Delta z_{max}$ in the tested $\Delta z_{max}$ range. We also note that changing $\Delta z_{max}$ mainly affects the lateral dimension, rather than the depth of resulting indents. The indentation typically creates dimples with depths less than or comparable to the hBN flake thickness even for the largest indents tested (Figure S2, Supplementary Information). This implies that the developed indentation procedure "pokes" hBN without deforming or damaging the substrate.

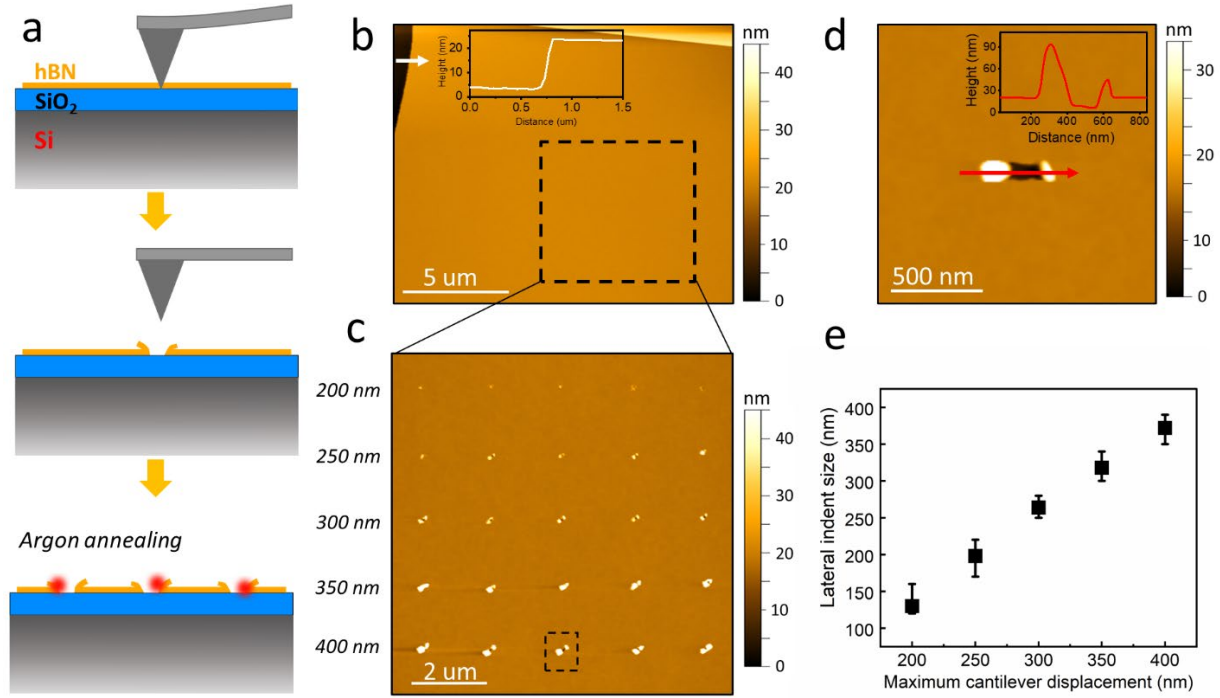

**Figure 1.** a) A schematic showing the activation of SPEs in hBN with the AFM nanoindentation technique followed by high temperature annealing in argon. b) An AFM image showing a thin hBN flake used for demonstrating the nanoindentation process. Upper inset: a height profile collected at the flake edge along the white arrow, showing a flake thickness of 19.8 nm. c) an AFM image of the area inside the dashed box in b) after nanoindentation. On the left lists the values of maximum cantilever displacement ($\Delta z_{max}$) during the indentation for each row of indents. d) An enlarged view of one specific indent inside the black dashed box in c), showing a dimple in the flake as well as two broken hBN areas next to the dimple. Note that the image was scanned at 45° with respect to the image in c) to present the indent along the horizontal direction. Inset: a height profile collected across the indent along the red arrow. e) Dependence of the lateral indent size as a function of the maximum cantilever displacement $\Delta z_{max}$.

Our findings are in contrast with reported AFM indentation experiments on TMDCs[11,18], where TMDC layers are completely deformed into indented dips with depths depending on $\Delta z_{max}$, without being torn apart by cantilevers. One reason can be attributed to a relatively rigid substrate ($SiO_2$-on-Si) that is not amenable to deformation, while in ref [11] a soft polymer substrate is used to facilitate the downward deformation of the TMDC layer. The difference between hBN and TMDCs in terms of their mechanical properties could also play an important role. High-quality hBN thin films around 15 nm thick are reported to have a Young's Modulus of 1.2 TPa[19], which is more than 4 times larger than the Young's Modulus of TMDCs[20]. This indicates that hBN is a stiffer material than TMDCs and tends to break more easily before significant deformation. An additional factor is the different types of cantilevers used among studies in terms of coating material, tip radius, etc., which might lead to varied cantilever-sample interactions and sample responses. Here, the indentation of hBN flakes without notable damage into substrates is, in fact, of unique advantage in practical applications where hBN flakes are placed on photonic components such as waveguides, resonators, cavities, etc.

After the nanoindentation, an annealing step is essential to activate SPEs in hexagonal boron nitride and meanwhile remove possible organic residues from the exfoliation process. In this work, all samples are annealed in argon at 1000 °C for 30 minutes to achieve efficient SPE activation.[7] In the following experiments, two types of AFM indented hBN samples are annealed and characterized: one with varied indentation sizes on the same flake to investigate the correlation between indent sizes and SPE generation, and another with a single indent size combining a high SPE yield and indentation precision. Figure 2a shows an hBN flake that has been indented with three different $\Delta z_{max}$, namely, 250 nm (group 1), 500 nm (group 2) and 1500 nm (group 3). Here, relatively large step sizes in $\Delta z_{max}$ between groups are used compared to the demonstration in Figure 1, which helps us to extract the dependence of SPEs on the lateral indent size in a broader size range. The lateral size distribution of all indents can be found in Figure S3 (Supplementary Information). Average indent sizes of 200 nm, 400 nm and 750 nm are measured in the three groups above. It needs to be noted that an indent size of 750 nm for $\Delta z_{max}$ = 1500 nm no longer fits into the linear relationship shown in Figure 1e. However, fitting for indents larger than 500 nm is out of the scope of this work since smaller indents are preferred in practice. We also note that even for $\Delta z_{max}$ as large as 1500 nm, the depth of resulting indents is still comparable to the hBN flake thickness (Figure S3d), again confirming that our technique has negligible impact on substrates. Overall, our indentation procedure can produce a designated indent size with decent repeatability. To further narrow down the size distribution of indents created by a given $\Delta z_{max}$, one can explore the rich parameters in the nanoindentation procedure, such as the indentation trigger point, cantilever extension/retraction velocity, dwell time at $\Delta z_{max}$, etc.

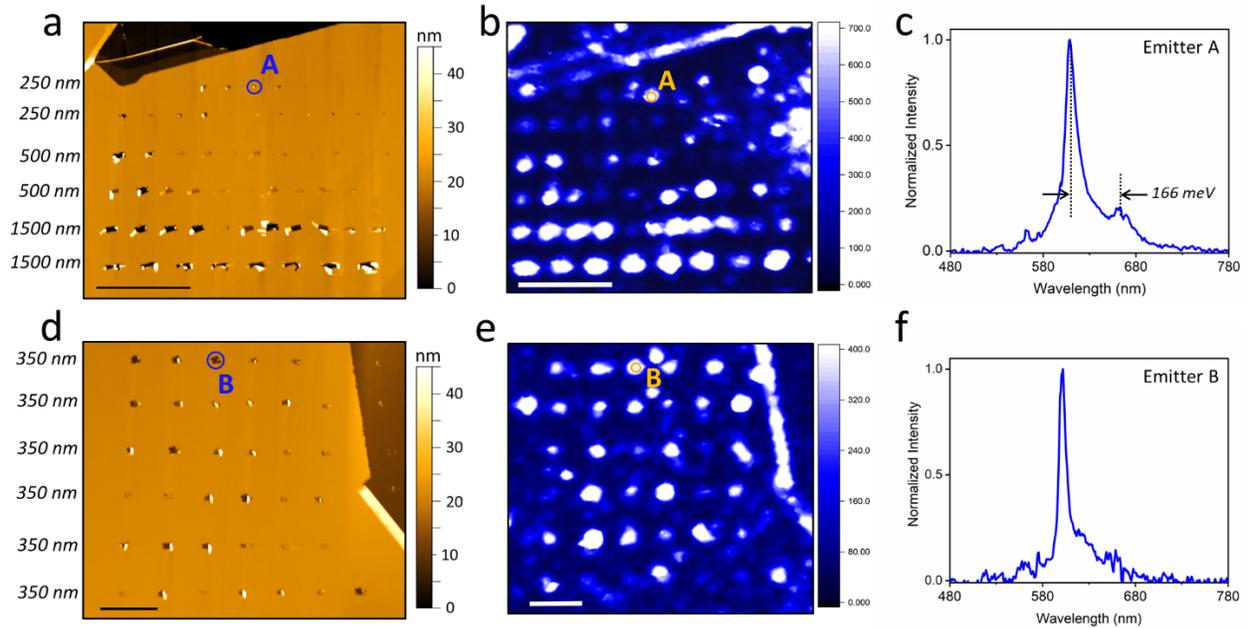

**Figure 2.** a) AFM image of the first flake after nanoindentation. Six rows of indents are created with the values of $\Delta z_{max}$ indicated on the left of each row. From top to bottom, the number of indents in each row is 5, 11, 11, 9, 10 and 8, respectively. Scale bar: 5 um. b) Photoluminescence (PL) map from the same area shown in a) after Ar annealing. Scale bar: 5 um. c) Background-subtracted spectrum from a representative emitter A circled in a) and b). A phonon sideband (PSB) that is 166 meV away from the zero-phonon line (ZPL) at 608 nm is identified. d) AFM image of the second flake after nanoindentation. Six rows of indents are created with the same $\Delta z_{max}$ of 350 nm. From top to bottom, the number of indents in each row is 5, 6, 6, 6, 6 and 7, respectively. Scale bar: 2 um. e) PL map of the same area shown in d) after Ar annealing. Scale bar: 2um. f) Background-subtracted spectrum from a representative emitter B circled in d) and e), showing a sharp ZPL line centered at 602 nm.

After annealing in Ar, the flake is scanned with our home-built confocal microscope at room temperature. A 532-nm-wavelength continuous wave (CW) pump laser is used for scanning at a laser power of 800 uW and the photoluminescence (PL) map from the AFM-indented area is collected (Figure 2b). Bright, isolated spots are identified at positions corresponding to the indented areas, with low background emission rates away from these spots. Spectroscopic measurements reveal that PL spectra from those bright spots typically show well-defined emission peaks on top of a low background. The spectrum of a representative emitter A is presented in Figure 2c. There is a sharp emission peak at 608 nm (2.04 eV) that matches the one of the reported zero-phonon lines (ZPLs) of hBN emitters (602 nm).[7] A broader peak that is ~166 meV away from the ZPL is recognized as a phonon sideband (PSB) corresponding to the in-plane optical phonons of hBN [7,21–23]. This confirms that our technique is capable to activate room-temperature emitters in hBN at indents as small as 200 nm in size.

Another important observation is that the yield of bright emission spots increases monotonically with the indent size (Figure 2b). While a high emitter yield would be desired, it is also preferable to work with emitters from relatively small indents such that the integration of emitters with photonic structures could be achieved with better precision[24–26]. To balance the above two factors, we nano-indented another flake with a maximum cantilever displacement of 350 nm. Advantages of this specific $\Delta z_{max}$ will be discussed later. The developed nanoindentation procedure has reproduced arrays of indents with comparable sizes (Figure 2d and Supplementary Information, Figure S4). An average indent size of 290 nm was measured along the lateral (horizontal) direction. PL map of this area (Figure 2e) shows arrays of bright emission spots that match well with the AFM indentation pattern. The PL spectrum of a typical emitter noted as emitter B in the area (Figure 2f) has a sharp ZPL emission centered at 602 nm (2.06 eV).

After confirming the existence of hBN emitters at nano-indented sites on both flakes above, we next evaluated their single-photon purity using a Hanbury Brown and Twiss (HBT) setup that measures the second-order autocorrelation functions g²(t) of emitter A and B (Figure 3a). Both emitters have g²(0) < 0.5 at zero-delay time, confirming that they are single photon sources. The experimental g²(t) data measured in our experiments can be well fitted with a three-level model expressed as:[5,27]

$$g^2(t) = 1 - (1+a) \cdot e^{-\frac{|t|}{\tau_1}} + a \cdot e^{-\frac{|t|}{\tau_2}}$$

where a is the photon bunching amplitude, while $\tau_1$ and $\tau_2$ are lifetimes of the excited and metastable states, respectively. After careful examination, 31 SPEs in total were identified out of all indented sites from the two hBN flakes above (Figure S5, Supplementary Information). The radiative decay lifetimes of all SPEs extracted from g²(t) range from 1.17 ns to 7.88 ns with an average lifetime of 3.86 ns (Figure 3b). Our observation agrees with the typical lifetimes reported for room-temperature SPEs in hBN.[7,28] To measure the brightness of these SPEs, PL intensities at a series of incident laser powers are recorded and then fitted using a first-order saturation model: $I = I_{sat}P/(P + P_{sat})$, where P($P_{sat}$) and I($I_{sat}$) are the incident (saturated) power and PL intensity, respectively. Figure 3c shows the fitted saturation curve of one of the brightest emitters that saturates at 4.1 mW with a saturation intensity of 1.1 Mcps, comparable to the brightest hBN emitters created by other techniques.[7,13,14,29] SPEs in this work typically exhibit linearly polarized emission (Figure 3c inset), indicating an atomic defect with in-plane dipole moment.[28] The photostability of SPEs is examined by recording their PL time traces when excited by the 532 nm laser at 800 uW. It is observed that roughly 50% of the SPEs maintain stable emission during the optical characterization, while 40% and 10% exhibit blinking and photobleaching behaviors, respectively (Figure S6, Supplementary Information). Blinking has been commonly observed in hBN SPEs[12,21,30] as well as in other types of solid-state SPEs.[31–33] It features switches between an "on" and "off" state with switching rates varying from one SPE to another. One important cause of blinking is the existence of other defects near the target SPE, which modifies the SPE charge state from time to time. According to previous reports, blinking is quite common for hBN SPEs created by deterministic techniques,[8,12,13] even after high-temperature annealing which is proved to improve emission stability.[29] This is likely related to the complex local environment near SPEs after the hBN lattice structures get modified deterministically by strain, electron/ion irradiation, etc. Strategies to effectively improve the photostability of deterministically formed hBN SPEs are therefore highly desirable.

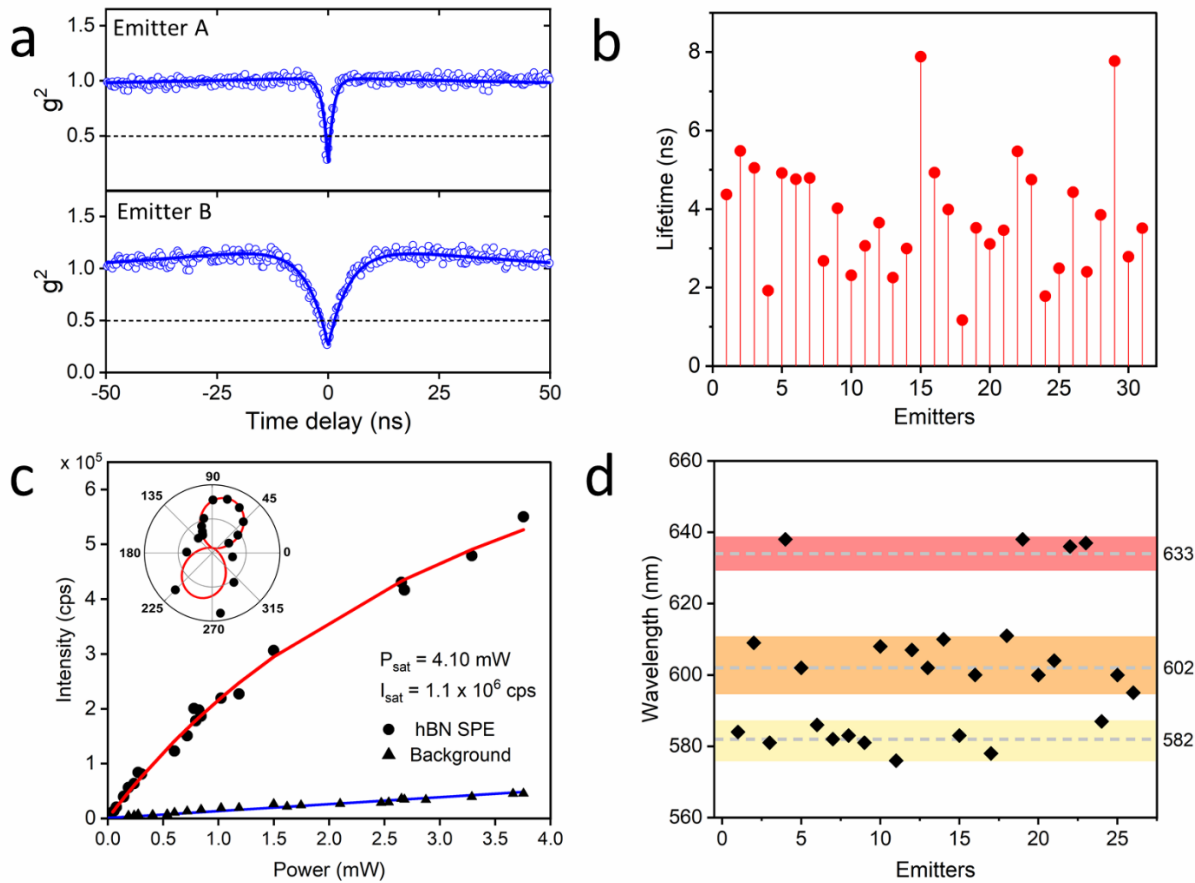

**Figure 3.** Photophysical properties of hBN emitters. a) Second-order autocorrelation measurements of emitter A & B. Blue circles are experimental data while blue solid lines are fitted curves using the three-level model. Both emitters show $g^2(0) < 0.5$ indicative of single photon emission. b) Distribution of radiative lifetimes from all 31 hBN SPEs, extracted from $g^2(t)$ fitting. The numbering of SPEs can be found in Figure S5 (Supplementary Information). c) Fluorescence saturation curve of a typical SPE with a saturation count of 1.1 Mcps. Inset: Emission polarization plot of a hBN SPE. d) Distribution of ZPLs for 26 photostable SPEs, revealing three classes of emitters with ZPLs at $582 \pm 6$ nm, $602 \pm 9$ nm and $633 \pm 5$ nm, respectively. The shaded bands are guides to the eye.

Figure 3d summarizes the distribution of ZPLs for all photostable SPEs from AFM-indented sites. The emitters can be classified into three groups with ZPLs centered at $582 \pm 6$ nm, $602 \pm 9$ nm and $633 \pm 5$ nm, all of which match the previously reported ZPLs of hBN SPEs in the visible spectral range.[7] One interesting fact is that, unlike SPEs created by other top-down techniques where a dominating portion of emitters show ZPLs at wavelengths shorter than 590 nm,[8,13] near 50% of emitters created with our technique exhibit ZPLs around 602 nm and another 40% show emission centered around 582 nm. 35% of the SPEs show PSBs in their spectra that are 167 ± 10 meV away from their corresponding ZPLs (Figure S7, Supplementary Information), in agreement with previous reports.[34–37] The relatively narrow distribution of ZPLs observed here implies that our technique is promising for deterministically creating hBN SPEs with predictable emission wavelengths.

Aside from SPEs, we also found emitters that possess some degrees of antibunching with $0.5 < g^2(0) < 1$. Considering the low fluorescence background of our samples, such emitters are most likely composed of more than one SPEs within spatially unresolvable spots.[38] Spectra collected from these emitters typically show broader emission peaks or multi-peaked emission, confirming the existence of multiple SPEs that could not be individually addressed (Figure S8, Supplementary Information). To demonstrate the correlation between $g^2(0)$ and indent size of corresponding emitters, we summarized the distribution of $g^2(0)$ values of emitters showing antibunched emission ($g^2(0) < 1$) as a function of their lateral indent sizes. Emitters from both hBN flakes with average indent sizes of 200 nm, 290 nm, 400 nm and 740 nm, are included for statistics. SPEs ($g^2(0) < 0.5$) are found from all four groups of indents, while clustered emitters ($0.5 < g^2(0) < 1$) are only found from indents with sizes ≥ 290 nm (Figure 4). After dividing the number of emitters with the total number of indents in each group, we obtained the emitter yields of both clustered emitters and SPEs for different indent sizes (bottom panel in Figure 4). The yield of clustered emitters shows a monotonic increase with indent sizes, from 0% for 200 nm-sized indents to over 40% for an indent size of 740 nm. In contrast, the SPE yield first increases with indent sizes, from 25% for indents of 200 nm up to 36% for indents of 400 nm, then drops for indents larger than 400 nm. The SPE yields for indents of 290 nm (32%) and 400 nm (36%) are, to the best of our knowledge, among the highest values ever reported for hBN SPEs created by top-down methods including FIB and nanopillar arrays.[8,13] Specifically, indents around 290 nm combine a high SPE yield and relatively small feature size, making them promising for the deterministic integration of hBN SPEs with photonic/plasmonic structures; The effective generation of clustered emitters in larger indents (740 nm) could be useful as well, for example, for sensing applications.

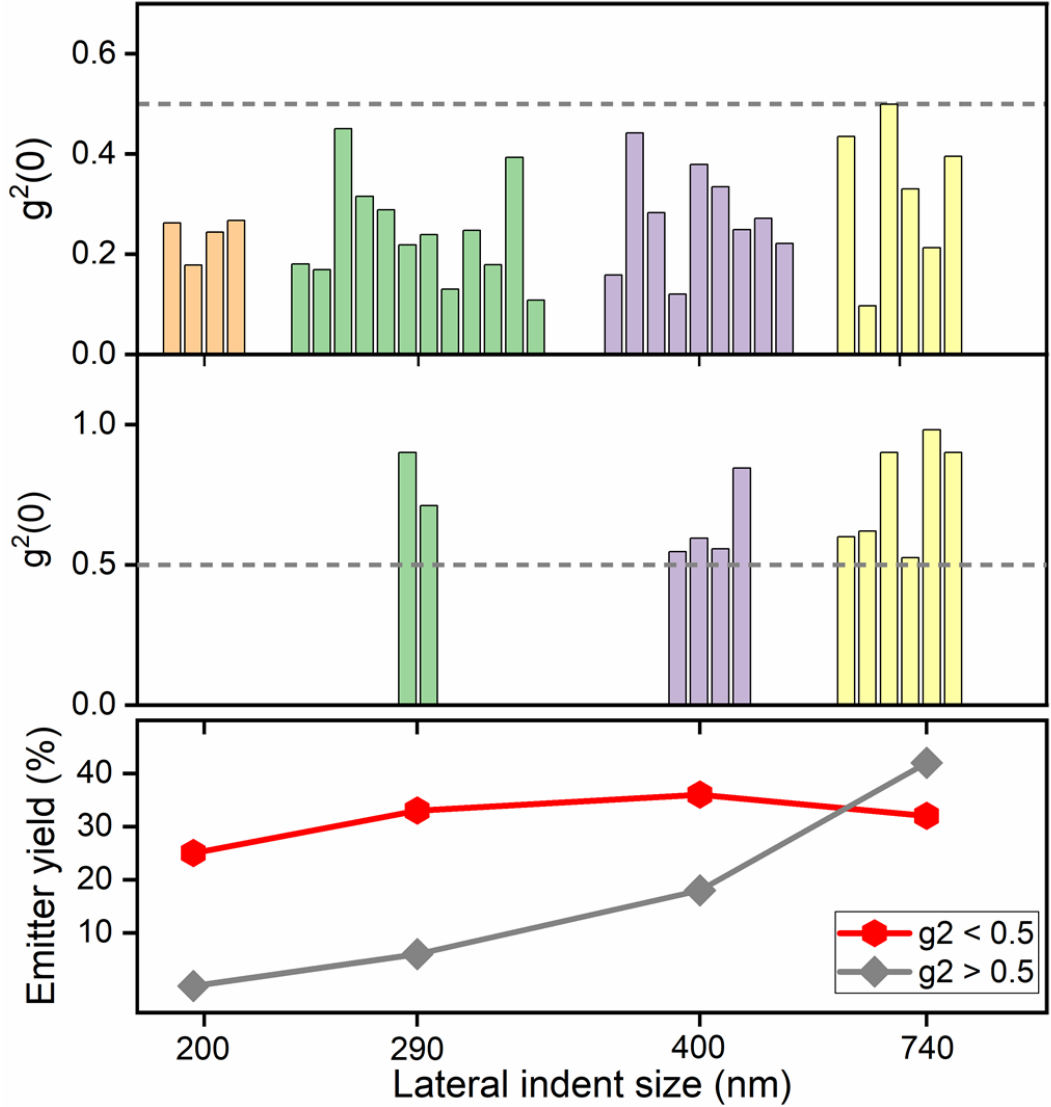

**Figure 4.** Top and middle panel: $g^2(0)$ values of SPEs ($g^2(0) < 0.5$, top) and clustered SPEs ($0.5 < g^2(0) < 1$, middle) as a function of corresponding indent sizes. Bottom panel: Yield of SPE and clustered emitters as a function of lateral indent size, calculated by dividing the number of emitters over the total number of indents created for each indent size.

We now discuss the formation mechanisms of hBN emitters created by AFM nanoindentation and their potential applications. First of all, the necessity of high-temperature annealing for SPE activation indicates that the emitters are most likely vacancy-related atomic defects as vacancy diffusion typically dominates at high temperatures.[39] To validate this, we performed a control experiment by annealing additional nano-indented hBN samples in argon at a lower temperature, namely, 800°C (see Supplementary Information for details). As seen in Figure S9, SPEs are found near the indents but with much lower emitter yields compared to hBN flakes annealed at 1000°C. This confirms that the formation of SPEs is less favorable at lower annealing temperatures likely due to the reduced vacancy diffusion rate. Our observation agrees with ref[7] where a monotonic increase in emitter yield is reported as a function of annealing temperature. However, further increasing the annealing temperature above 1000°C is not preferred in practice considering the degradation of involved materials. Other post-AFM treatments such as plasma etching could be explored as alternative ways to achieve higher SPE yields. As been reported previously, plasma treatments is capable of generating vacancies in hBN lattices as well as creating hBN SPEs.[29,40,41] Adding a plasma etching step after AFM nanoindentation might help to improve the SPE yield by increasing the probability of forming vacancy-related defects that act as SPEs near the indents.

The determination of hBN SPE locations with nanoscale precision has not be achieved here due to the diffraction-limited resolution of the confocal microscope. Nevertheless, we infer that SPEs are most likely formed either at the edge of indented dips or on the bent, delaminated hBN areas next to the dips. The former favors SPE formation as defects tend to accumulate at structure/grain boundaries,[13,42] while the latter could activate strain-induced SPEs due to the large curvature near the bent hBN areas. It is highly possible that both mechanisms play a role here, resulting in higher SPE yields than those reported for the at-the-boundary creation or strain engineering.[8,13] To locate SPEs with a higher resolution, one can use a near-field scanning optical microscope (NSOM),[43] scanning antenna microscope[44] or combined confocal microscope-AFM setup[45] to get correlated PL and topographical maps.

One of the most important advantages of hBN SPEs activated with AFM indentation is the combined high SPE yield and high-precision position control. As been demonstrated above, a SPE yield of 32% is obtained for indents less than 300 nm in dimension, which sets the new record for deterministic creation of hBN SPEs on non-structured substrates. It also needs to be emphasized that indents created by AFM features sharp, well-defined edges, in contrast to holes/spots induced by radiation methods that typically have poorly defined boundaries due to the diffuse nature of radiation beams. On the other hand, our technique can be extended to hBN flakes with various thicknesses by simply adjusting the indentation parameters, making it more versatile than strain engineering with structured substrates as the latter typically requires thin hBN films to achieve desired deformation.

The position-controlled creation of hBN SPEs with AFM nanoindentation could be a promising route for various quantum photonic applications. For example, it enables the efficient on-chip integration of hBN SPEs with photonic waveguides or cavities by creating SPEs at desired positions, in contrast to previously studies where hBN SPEs are coupled to photonic structures either randomly[25,46–48] or more deterministically, yet with extensive alignment efforts.[49] The developed technique also offers a great potential for coupling hBN SPEs to plasmonic nanostructures to achieve strong emission enhancement.[50] Compared to randomly activated SPEs in hBN that do not possess recognizable topographical features, SPEs near the nano-sized indents make it possible to deterministically assemble plasmonic cavity/antenna structures[51,52] with high SPE coupling efficiency. In addition, AFM-induced hBN SPEs could be used in other studies on hBN SPEs that might benefit from deterministic emitter coupling, such as sensing, emission tuning,[53,54] quantum non-linear optics,[55,56] and more.

**Conclusion**

We demonstrated a new route to deterministically create room-temperature SPEs in hBN utilizing AFM nanoindentation. The technique is applied to thin hBN flakes on an unpatterned $SiO_2$-coated Si substrate. By carefully controlling the indentation parameters, indents with lateral sizes ranging from 200 nm to over 700 nm are obtained, without notable damage or deformation of the substrate. After high temperature annealing in argon, hBN SPEs are activated near the indents for various indent sizes. The SPEs show a relatively narrow distribution of emission wavelengths, with over 80% of them emitting around 583 nm and 602 nm. A maximum SPE yield of 36% is obtained for an indent size of 400 nm, and smaller indents around 290 nm give a comparably high SPE yield. Our method involves no lithographic or other patterning steps, so that fab-induced fluorescence contamination is avoided. While the nature of these indentation induced SPEs has to be studied in more detail, it is inferred that the creation of structural edges and highly strained hBN areas near the indents could be responsible for the observed high SPE yield. The efficient SPE activation on flat, chip-compatible substrates with high precision allows controlled coupling of hBN SPEs with plasmonic and photonic devices for various quantum information applications. Hence, our results open exciting avenues towards the on-chip deterministic integration of hBN SPEs and future on-chip quantum photonic devices.

**Materials and Methods**

*Sample Preparation*

hBN flakes were obtained using a standard mechanical exfoliation method. High-quality hBN crystals produced by high-pressure synthesis were picked up by a sticky tape and then exfoliated to new tapes multiple times to get thin flakes. After several exfoliations, thin hBN flakes were transferred to flat silicon substrates coated with a 285 nm-thick $SiO_2$ layer. The substrates were cleaned with acetone, methanol, isopropanol and de-ionized water both before and after the transfer procedure. After transferring, an optical microscope was used to identify and locate hBN thin flakes based on the thickness-dependent reflection colors of hBN. The exact thicknesses of those thin flakes were then measured by an atomic force microscope (Cypher S AFM, Asylum Research). Only smooth flakes (without folds or wrinkles) with thicknesses less than 25 nm were selected for the rest of experiments.

Nanoindentation on pre-selected hBN flakes was carried out with the same AFM operated in the contact mode. Cantilevers with a diamond-like carbon coating layer (Tap300DLC, 300 kHz, 40N/m, BudgetSensors) were used. The indentation was triggered after the cantilever got in contact with the sample surface and the deflection (measured on the photo-detector, in the unite of voltage) reached 0.3 V. The maximum cantilever displacement along the vertical direction was adjustable from 0 nm up to 1500 nm to create indents with different lateral sizes. A displacement velocity of 100 nm/s was used for both the extension and retraction of cantilevers, with a dwell time of 2 s at the maximum displacement position. After arrays of indents were generated, the AFM was switched to tapping mode to get topographical images of the indented areas.

Argon annealing was performed in a tube furnace (Lindberg/Blue M) following the indentation experiments. The samples were heated from room temperature to 1000°C with a ramping time of ~40 mins. After being held at 1000°C for 30 mins in a continuous argon gas flow, the samples were allowed slowly cool down to room temperature in 6 hours.

*Optical Characterization*

Optical characterization was performed by a home-built scanning confocal microscopy setup based on a commercially available inverted microscope body (Nikon Ti-U) with the objective scanned by a P-561 PIMars piezo stage with an E-712 controller and Alignment Firmware (Physik Instrumente). Optical pumping for continuous wave (CW) excitation experiments was provided by a 200 mW continuous wave 532 nm laser (RGB Photonics). The fluorescence to be detected was separated from the pump radiation by a main dichroic beamsplitter (550 nm long-pass DMLP550L, Thorlabs), and subsequently a 550 nm long-pass filter (FEL0550, Thorlabs). The collected fluorescence was passed through a 100 um pinhole. Two single-photon avalanche detectors (SPADs) with a 30 ps time resolution and 35% quantum efficiency at 650 nm (PDM, Micro-Photon Devices) were used for autocorrelation measurements. An SPAD with 69% quantum efficiency at 650 nm (SPCM-AQRH, Excelitas) was used for scanning and saturation measurements. Spectral measurements were performed with a QE65000 visible-to-near infrared spectrometer (Ocean Insight). To measure the emission polarization, a broadband analyzer (LPVISC050-MP2, Thorlabs) was placed in the optical collection path and rotated in 20° increments. A time trace of the emission intensity was recorded, and averaged photon counts were correlated with analyzer angles.

## Acknowledgements

The authors acknowledge S. I. Bogdanov, A. Senichev and A. B. Solanki for helpful discussions on the optical characterization of hBN SPEs. This work is supported by the U.S. Department of Energy (DOE), Office of Science through the Quantum Science Center (QSC), a National Quantum Information Science Research Center and National Science Foundation Award 2015025-ECCS.

# Supplementary Information

## Creating Quantum Emitters in Hexagonal Boron Nitride Deterministically on Chip-Compatible Substrates

*Xiaohui Xu[1], Zachariah O. Martin[2], Demid Sychev[2], Alexei S. Lagutchev[2], Yong Chen[2,3,4], Takashi Taniguch[5], Kenji Watanabe[5], Vladimir M. Shalaev[2], Alexandra Boltasseva[1,2]*

[1]School of Materials Engineering, Purdue University, USA

[2]School of Electrical and Computer Engineering, Purdue University, USA

[3]Department of Physics and Astronomy, Purdue University, USA

[4]Department of Physics and Astronomy, Aarhus University, Denmark

[5]National Institute for Materials Science, Japan

# Contents

I. Measurement and determination of lateral indent size

Here we specify the method used to measure the lateral size of AFM-induced indents in this work. It is known that the AFM probes create various types of artifacts in AFM scanning/imaging.[1] Specifically, when scanning a feature that is lower than the surface, such as an indent as in this work, the lateral size of the indent can appear smaller than its actual size due to the width of the AFM probe.[2] The effect is illustrated in Figure S1a. Therefore, rather than taking the length of the recessed portion of the AFM height profile as the indent size, we denote the lateral indent size to be the distance between the two highest features across the indent, as shown in Figure S1b.

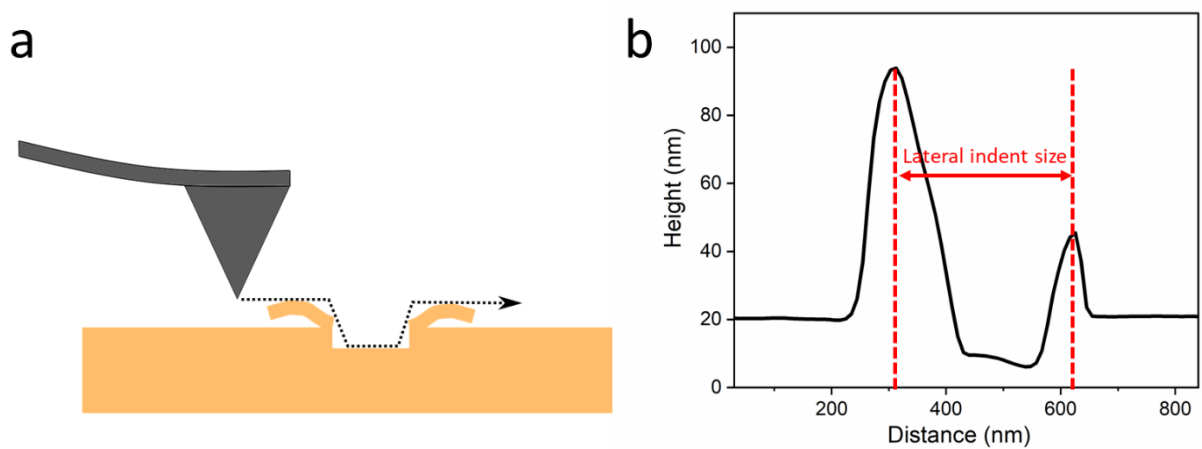

**Figure S1.** a) Schematic showing the effect of AFM cantilever when scanning an indent structure in hBN. Due to the width of the AFM tip, the indent profile cannot be accurately reproduced. b) Definition of the lateral indent size in this work, which refers to the distance between the two highest points across the indent. The height profile is taken from Figure 1d from the main text as an example.

II. Dependence of the indentation depth on the maximum cantilever displacement

Figure S2a shows an AFM image of a hBN flake after indentation. The indents were created with values of the maximum cantilever displacement along the vertical direction ($\Delta z_{max}$) indicated on the left of each indent row. The flake has a thickness of ~11.5 nm as measured on the flake edge (Figure S2b). To show the indentation depth induced by different $\Delta z_{max}$, the height profile across one representative indent was collected from each row of indents. As illustrated in Figure

S2c, while the lateral indent size increases monotonically with $\Delta z_{max}$, the indentation depth shows little dependence on $\Delta z_{max}$, with magnitudes comparable or less than the flake thickness for all $\Delta z_{max}$ values tested. Our results indicate that the indentation procedure used in this study has negligible effect or damage on the substrate underneath hBN.

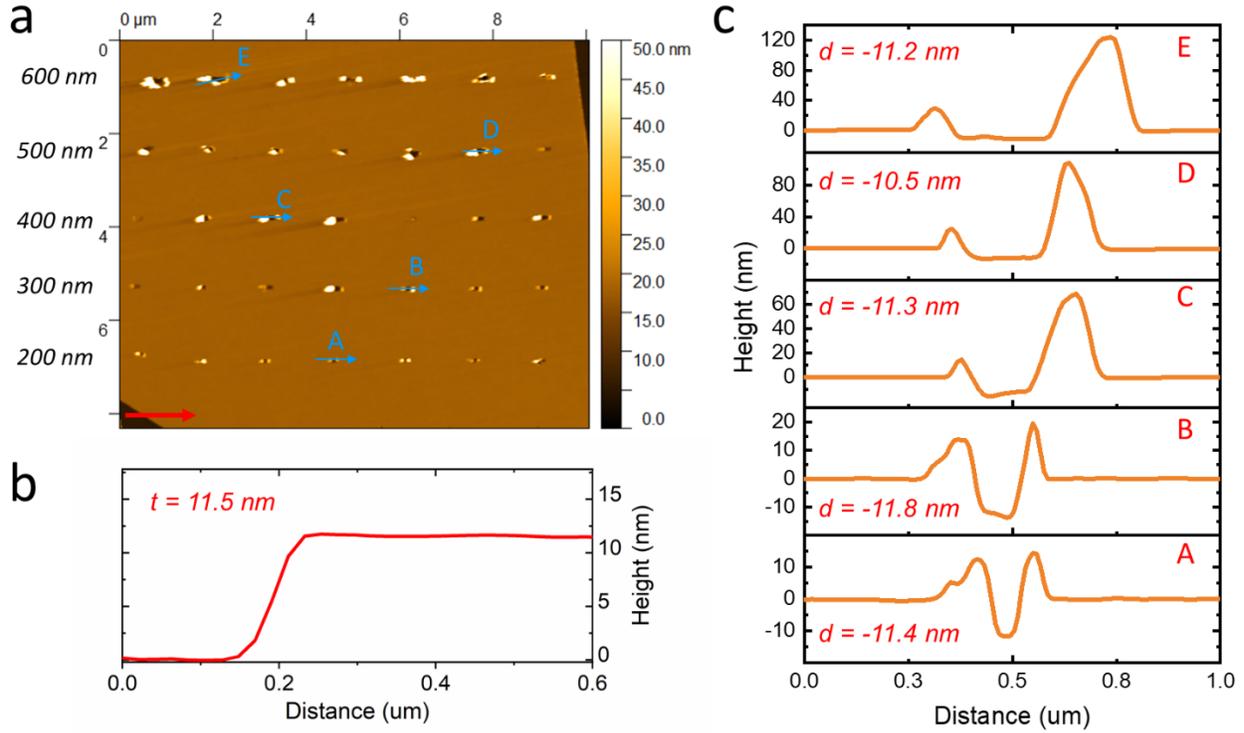

**Figure S2.** a) AFM image of an hBN flake after nanoindentation. Indents in the same row are created with the same maximum cantilever displacement $\Delta z_{max}$ as indicated on the left. b) Height profile along the solid red arrow in a), showing a flake thickness t = 11.5 nm. c) Height profiles across the indents noted as A—E along the blue arrows in a). Value of d noted for each indent in the plot indicates the measured indentation depth.

## III. Additional data from hBN flakes used for optical measurement

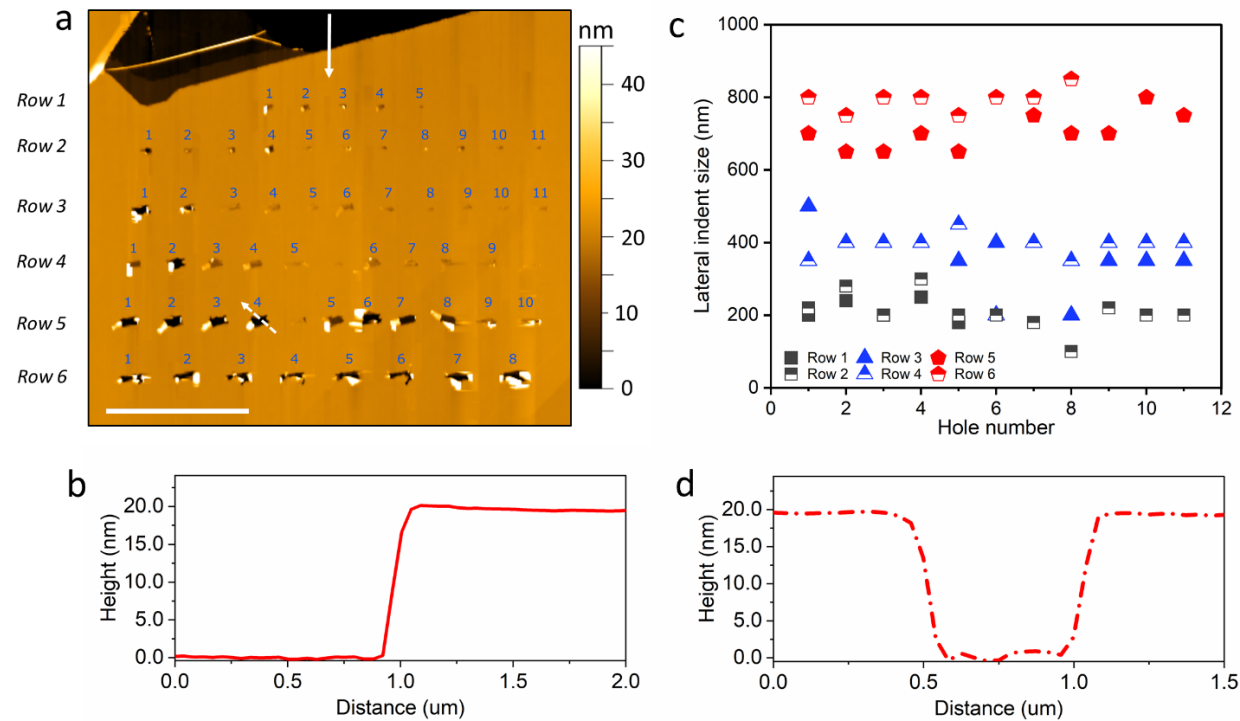

**Figure S3.** a) AFM image of the same flake shown in Figure 2a, with all indents numbered for reference. Scale bar: 5 um. b) Height profile at the flake edge along the solid white arrow in a), showing a flake thickness of 19.5 nm. c) Distribution of the lateral indent size for all indents in a). Note: the indent sizes are measured along the lateral (horizontal) direction. d) The height profile of one of the largest indents (in terms of lateral size), i.e., indent number 4 in row 5, along the white dashed arrow, showing an indent depth comparable with the flake thickness.

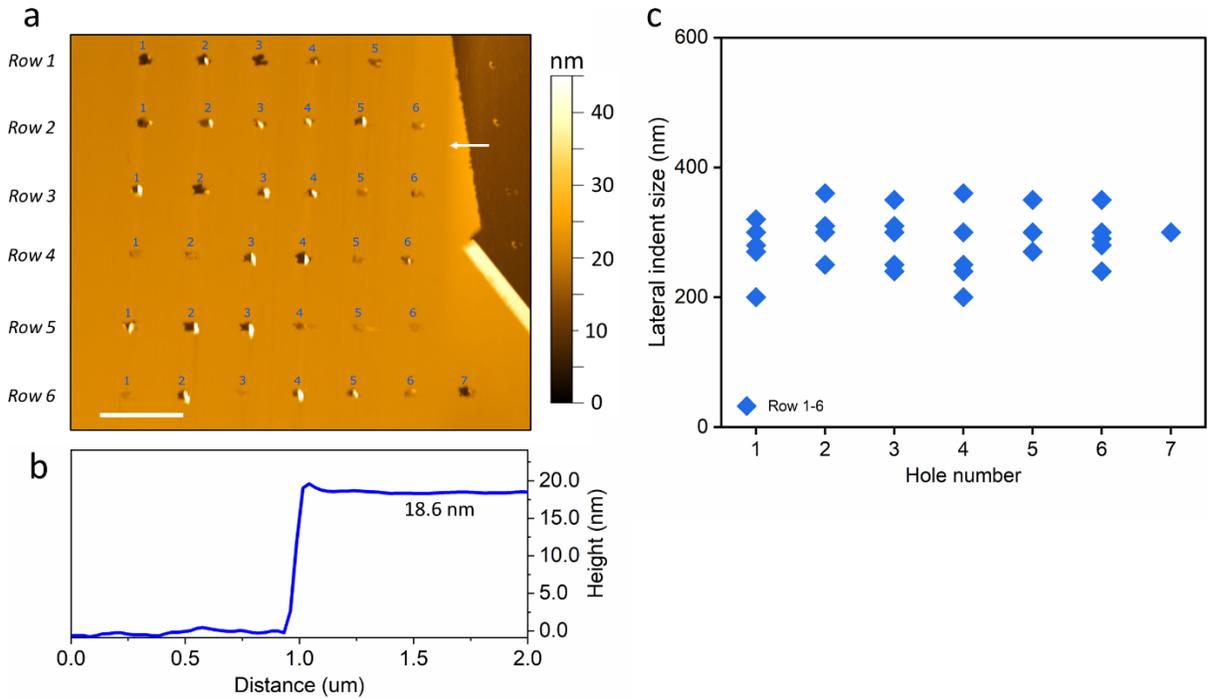

**Figure S4.** a) AFM image of the same flake shown in Figure 2d, with all indents numbered for reference. Scale bar: 2 um. b) Height profile at the flake edge along the solid white arrow in a), showing a flake thickness of 18.6 nm. c) Distribution of the lateral indent size for all indents in a).

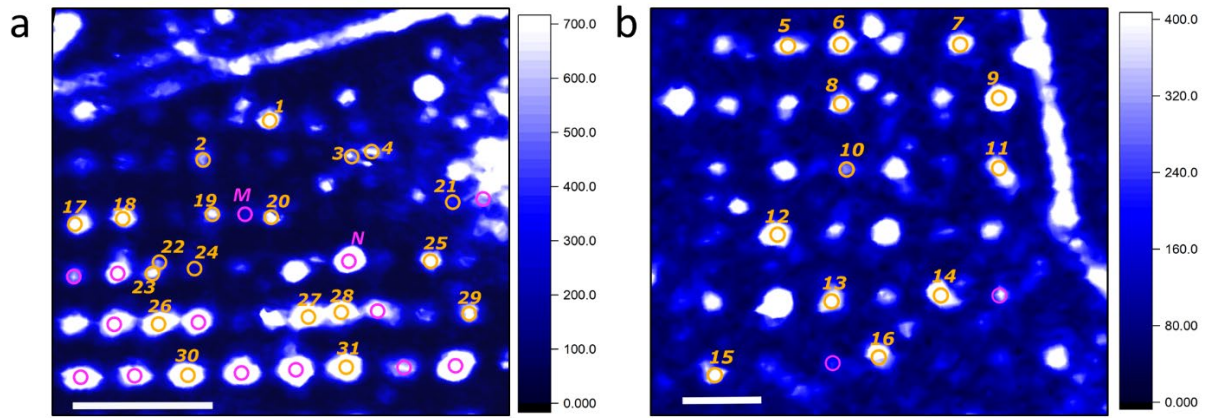

**Figure S5.** Notation of all emitters found on the two hBN flakes shown in Figure 2b & 2e. SPEs are circled and numbered in yellow, while clustered emitters are marked with magenta circles.

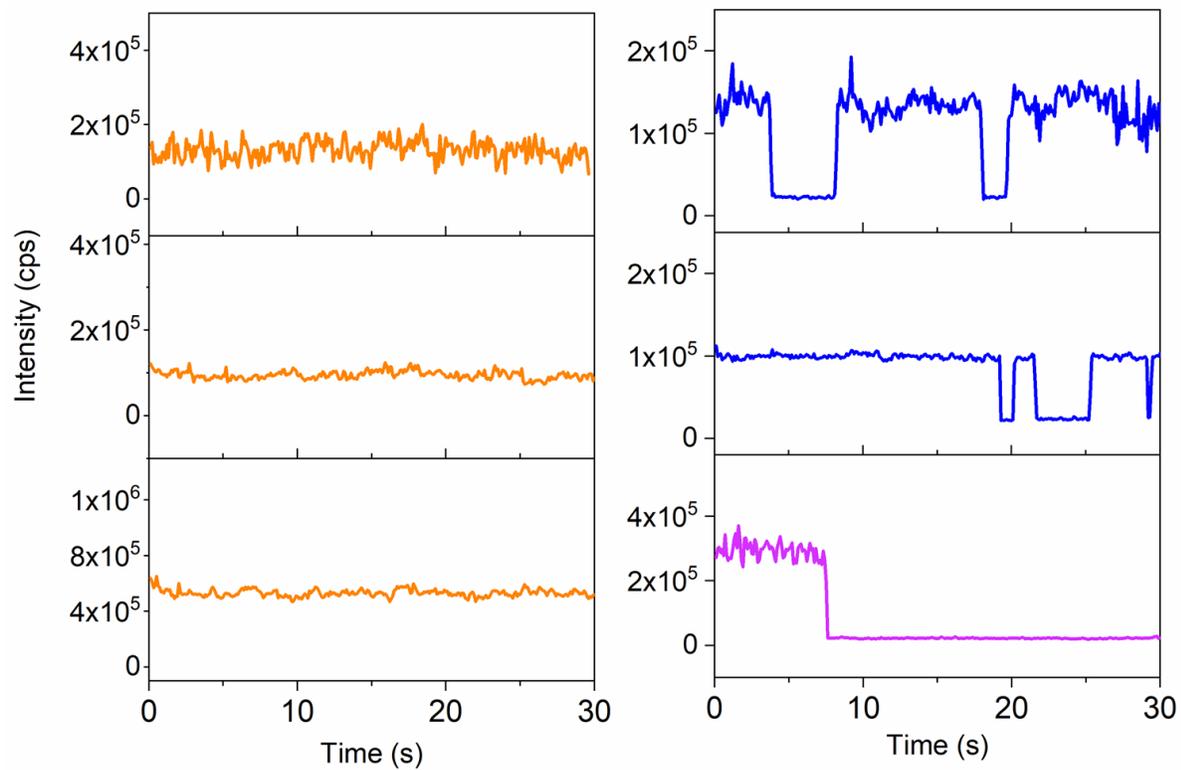

**Figure S6.** Stability of several representative SPEs showing photostable (three orange curves), blinking (two blue curves) and photobleached (one purple curve) emitters.

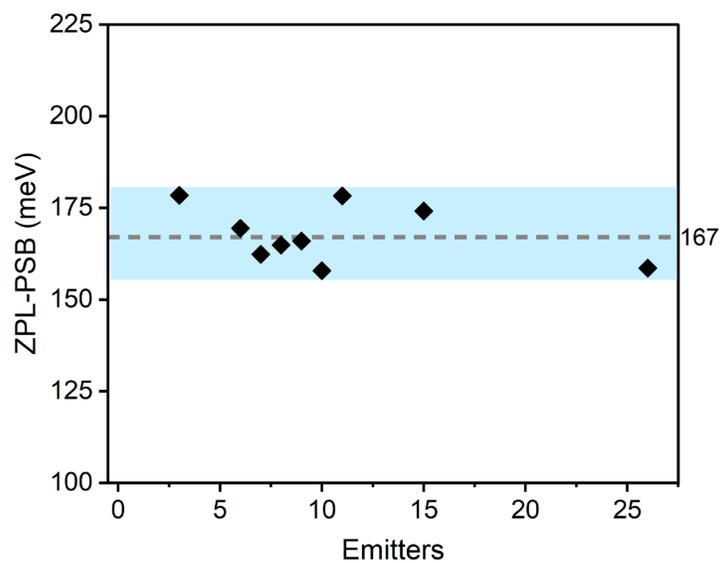

**Figure S7.** Difference in the energy of ZPL and PSB for SPEs that show well-defined PSBs in spectra. The numbering of emitters is consistent with Figure 3b and Figure S4. The shaded band is a guide to the eye spanning over $167 \pm 10$ meV.

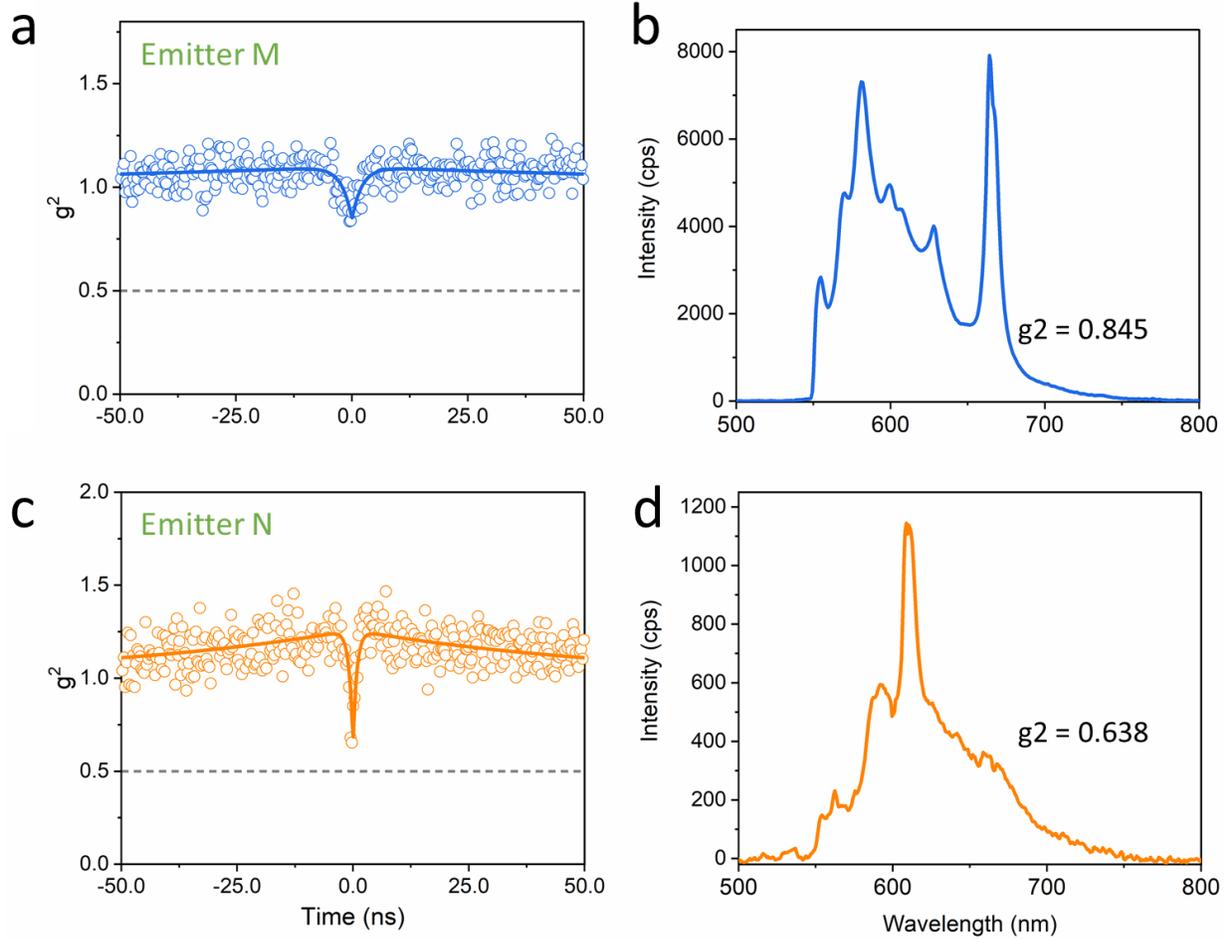

**Figure S8.** a) and c) Second-order autocorrelation measurements of two clustered emitters M & N, both showing $0.5 < g2(0) < 1$. Locations of emitter M & N are noted in Figure S4. c) and d) Corresponding spectra of Emitter M & N showing multiple emission peaks and broad emission, indicative of the existence of more than one SPEs per site.

IV. **Control experiment with hBN flakes annealed at 800 °C**

To investigate the effect of annealing temperature on the hBN emitter yield, we prepared additional hBN flakes, indented with AFM, and annealed them in argon at a lower temperature (800 °C) for 30 mins. Figure S9a shows one hBN flake with arrays of indents that were all created with $\Delta z_{max}$ = 350 nm, the same $\Delta z_{max}$ that leads to a SPE yield of 32% at an annealing temperature of 1000 °C. Here, the hBN flake has a thickness around 20 nm (Figure S9c) that is comparable to hBN flakes used in the main text. After argon annealing, the sample was again characterized with the confocal microscope to identify stable emitters. Figure S9b is a PL map of the indented area, with emitters identified in solid and dashed circles corresponding to SPEs and clustered emitters, respectively. SPEs found in this area share similar photophysical properties with those obtained by annealing at 1000 °C in terms of emission wavelength (S9d & S9e). However, the SPE yield is calculated to be 12.8%, much lower than that on hBN flakes annealed at 1000 °C (32%). Our observation above supports the hypothesis that atomic defects giving single photon emission at indented sites on hBN are vacancy-related structures: as the annealing temperature increases, the diffusion of vacancies is promoted, thus leading to a higher probability of forming vacancy-based defects that act as SPEs.

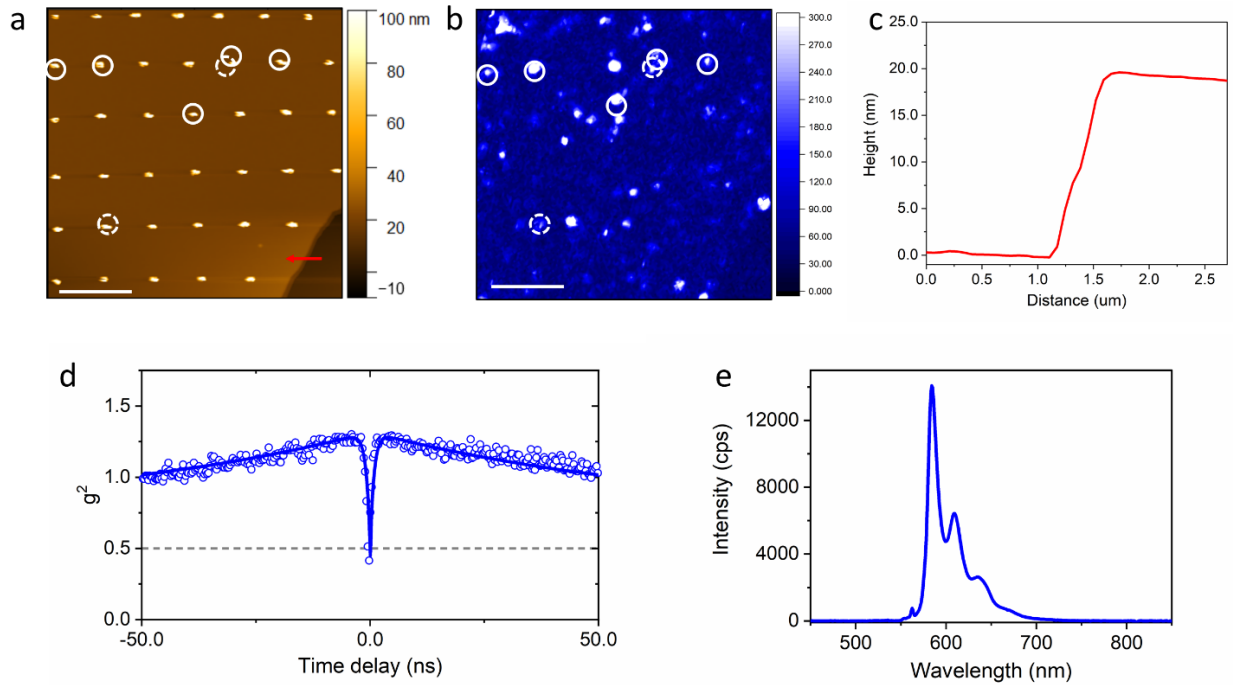

**Figure S9.** a) AFM image of an hBN flakes after nanoindentation. The flake was indented with the maximum cantilever displacement $\Delta z_{max}$ = 350 nm. Scale bar: 2 um. b) PL maps of the same indented areas shown in a) after being annealed in argon at 800 °C. SPEs and clustered emitters are found and marked with while solid and dashed circles, respectively. Scale bar: 2 um. The corresponding indents are also circled out in a). c) Height profile at the flake edge along the red arrow marked in a), showing a flake thickness of ~20.0 nm. d) Second-order autocorrelation measurement and fitting of a SPE from the area shown in a), The blue circles are experimental data, while the solid blue line is the fitting curve with the three-level model. e) Background-subtracted spectrum of the same SPE, showing a ZPL at 581 nm and a PSB at the longer wavelength.